\documentstyle[preprint,aps]{revtex}

\begin{document}

\draft


\title{Toward a Complete Analysis of the Global Structure\
of Kerr-Newman Spacetime}

\author{Hongsu Kim}

\address{Department of Physics\\
Ewha Women's University, Seoul 120-750, KOREA}

\date{November, 1996}

\maketitle

\begin{abstract}
An attempt is made to supplement Carter's partial investigation
of the global structure of Kerr-Newman spacetime on the symmetry
axis. Namely, the global structures of $\theta =$ const.
timelike submanifolds of Kerr-Newman metric starting from the
symmetry axis all the way down to the equatorial plane
are studied by introducing
a new time coordinate slightly different from the
usual Boyer-Lindquist time coordinate. It turns out
that the maximal analytic extension of $\theta =\theta_{0}$
($0 \leq \theta_{0} < \pi/2$) submanifolds is the same as that of 
the symmetry axis first studied by Carter
whereas $\theta = \pi/2$ equatorial plane has 
the topology identical to that of the Reissner-Nordstrom spacetime.
General applicability of this method to Kerr-Newman-type black
hole solutions in other gravity theories is discussed as well.
\end{abstract}


\narrowtext


{\bf I. Introduction}
\\
 Generally speaking, a complete analysis of the global structure of full, 
4-dim. spacetime solution of Einstein equations is a formidable task. In 
certain special cases, however, there are intrinsically singled-out 
timelike 2-dim.
submanifold, ``${\it T}$" of the full, 4-dim. manifold, ``${\it M}$" 
which can be simply
analyzed. For instance, in the Schwarzschild or Reissner-Nordstrom (RN)
solution$^{1}$, the spacetime has topology of ${\it M} = R^2 \times S^2$ 
because of 
the spherical symmetry and we can think of each point of the timelike 2-dim.
submanifold ${\it T} = R^2$ as representing a two-sphere $S^2$ whose 
radius is 
the $r$-value at that point. In these two spherically-symmetric cases, 
the global structures of the submanifolds $T$ exactly mirror those of the 
full 4-dim. 
manifold $M$. In the case of Kerr-Newman solution$^2$, however, this is not 
the case. There are incomplete, inextendable geodesics in the Kerr-Newman
spacetime which do not lie in any totally geodesic, timelike 2-dim. 
submanifold $T$. Despite this fact, by adopting the new coordinates 
introduced by Boyer-Lindquist$^3$ (which can be thought of as the 
generalization of the Schwarzschild coordinates), Carter$^3$ was able to 
demonstrate the global structure of a timelike 2-dim. submanifold of the 
Kerr-Newman solution. And the 2-dim. submanifold is the $(t, r)$-plane in
Boyer-Lindquist coordinates which represent the ``symmetry axis" 
($\theta = 0$) of the Kerr-Newman spacetime. Then by applying the methods 
of Finkelstein$^4$ and Kruskal$^5$, he obtained the maximal analytic 
extension 
(depicted in the Carter-Penrose diagram$^1$) of the geometry on the symmetry 
axis. He then went on to conjecture that the analytic extension of the full,
4-dim. Kerr-Newman manifold would have basically the same topology except
that the ``ring singularity" at $\Sigma = r^2 + a^2 \cos^2 \theta =0$$^1$
of the full spacetime cannot be represented in his 2-dim. picture, i.e.,
in the extended conformal diagram of the symmetry axis$^3$.
In the present work, we would like to supplement Carter's partial 
investigation of the global structure of Kerr-Newman spacetime.
Namely, we consider the geometry of the ``$\theta =$ const."
timelike submanifolds of Kerr-Newman spacetime including both the
``symmetry axis" ($\theta = 0$) and the ``equatorial plane"
($\theta = \pi/2$) and examine their global structures. Since the
geometry of $\theta = \theta_{0}$ (with $0 < \theta_{0}\leq \pi/2$)
submanifolds is essentially 3-dim. whereas that of the symmetry axis 
($\theta_{0} = 0$) is {\it effectively} 2-dim. (since it has a 
degeneracy along $\phi$ direction), things get more 
involved compared to the case of the symmetry axis discussed by Carter.
Therefore in order to make the analysis tractable, we shall introduce
a ``new time coordinate" $\tilde{t}$ slightly different from the
usual Boyer-Lindquist$^3$ time coordinate $t$ based on the philosophy
that the global structure remains unaffected under coordinate changes.
From there one can then apply the same methods of Finkelstein and Kruskal 
to obtain the
maximal analytic extensions of the conformal diagrams of the
$\theta = \theta_{0}$ submanifolds. Then one can readily see that the
extended conformal diagrams
turn out to be identical to the one obtained by Carter for the cases
$0 \leq \theta_{0} < \pi/2$ as has been anticipated. For the case 
$\theta_{0} = \pi/2$, i.e., the equatorial plane, however, one can see
that the 
conformal diagram turns out to be that of the RN spacetime$^{1,3}$ explicitly
showing the existence of the ring singularity $\Sigma =0$ (which of 
course becomes $r = 0$ on $\theta_{0} =\pi/2$ equatorial plane) owned by
the full, 4-dim. Kerr-Newman manifold.  As stated earlier, the peculiarity 
with the global structure of the symmetry axis is that the 2-dim. metric
of the symmetry axis is everywhere analytic and non-singular and hence the 
ring singularity $\Sigma = 0$ is absent there in its extended conformal
diagram although it does exist in full, 4-dim. manifold. As a result, on the
symmetry axis, one can pass through the ring singularity (or more precisely,
through $r = 0$) and extend to ``negative'' values of $r$. This possibility
causes some trouble since in this region containing the ring singularity,
there may exist closed timelike curves which lead to the causality
violation as pointed out by Carter$^3$. 
As we shall see shortly, the maximally extended conformal diagram of the
$\theta = \theta_{0}$ (with $0 < \theta_{0} < \pi/2$) submanifolds
of Kerr-Newman spacetime remain the same, i.e., still take the same
structure as that of the symmetry axis.
The extended conformal diagram of the equatorial plane, on the other hand, 
takes exactly the same structure as that of the RN solution. 
As a result, it does have the ring singularity at $\Sigma = 0$
(or more precisely at $r = 0$ since $\theta_{0} = \pi/2$) and one cannot, on
the equatorial plane, extend to negative values of $r$.
In short, the result of our analysis of the global structure of 
$\theta = \theta_{0}$ (with $0 < \theta_{0} \leq \pi/2$) submanifolds
supplements that of the global structure of the symmetry axis
studied by Carter to bring us a clearer overview of the global 
structure of the full Kerr-Newman spacetime. And particularly, our study
of the global topology of the equatorial plane confirms the existence of
the ring singularity and supports the general belief that the spacetime
produced by physically realistic collapse of even nonspherical bodies 
would be
qualitatively similar to the spherical case, i.e., the RN geometry.
\\
{\bf II. Global structure of $\theta =$const. submanifolds of 
Kerr-Newman geometry}
\\
Now consider the stationary, axisymmetric Kerr-Newman solution of the 
Einstein-Maxwell equations. The Kerr-Newman metric solution is given in 
Boyer-Lindquist coordinates as$^{1,3}$
\begin{eqnarray}
ds^2 = &-& [{\Delta - a^2 \sin^2 \theta \over \Sigma}] dt^2 - 
{2a\sin^2 \theta (r^2 + a^2 - \Delta) \over \Sigma} dt d\phi \nonumber \\
&+& [{(r^2 + a^2)^2 - \Delta  a^2 \sin^2 \theta \over \Sigma}]  \sin^2 \theta
d\phi^2 + {\Sigma \over \Delta} dr^2 + \Sigma d\theta^2  
\end{eqnarray}
where $\Sigma \equiv r^2 + a^2 \cos^2 \theta$ and $\Delta \equiv
r^2 - 2Mr + a^2 + e^2$ with $M$ being the mass, $a$ being the angular
momentum per unit mass and $e$ being the total $U(1)$ charge of the hole.
We are now particularly interested in the $\theta =$const. timelike surfaces
as {\it submanifolds} of this Kerr-Newman spacetime. 
Namely, consider the $\theta = \theta_{0}$ ($0 \leq \theta_{0} \leq 
\pi/2$) timelike submanifolds of Kerr-Newman spacetime with the metric
\begin{eqnarray}
ds^2 = &-& [{\Delta - a^2 \sin^2 \theta_{0} \over \Sigma}] dt^2 -
{2a\sin^2 \theta_{0}(r^2 + a^2 - \Delta) \over \Sigma} dt d\phi \nonumber \\
&+& [{(r^2 + a^2)^2 - \Delta a^2 \sin^2 \theta_{0} \over \Sigma}]  \sin^2 
\theta_{0}
d\phi^2 + {\Sigma \over \Delta} dr^2 
\end{eqnarray}
where $\Sigma = r^2 + a^2 \cos^2 \theta_{0}$ now. 
These $\theta = \theta_{0}$ surfaces have metrics which are
literally 3-dim. in structure and possess an off-diagonal component in an
intricate way. Thus in order to make the study of global structure of
$\theta = \theta_{0}$ submanifolds tractable, here we consider a coordinate
transformation which is defined {\it only} on the $\theta =$const. timelike
submanifolds. Namely consider a transformation from the Boyer-Lindquist
time coordinate ``$t$'' to a new time coordinate ``$\tilde{t}$'' given by
\begin{eqnarray}
\tilde{t} = t - (a\sin^2 \theta_{0}) \phi
\end{eqnarray}
with other spatial coordinates $(r, ~\phi)$ remaining unchanged.
Note that the new time coordinate $\tilde{t}$ is different from the old one
$t$ only for ``rotating'' case ($a \neq 0$) and even then only for 
$\theta_{0} \neq 0$. In terms of this ``new'' time coordinate $\tilde{t}$,
the metric of $\theta = \theta_{0}$ submanifolds becomes
\begin{eqnarray}
ds^2 &=& - {\Delta \over \Sigma} d\tilde{t}^2 + {\Sigma \over \Delta} dr^2 +
\Sigma  \sin^2 \theta_{0} (-{a\over \Sigma}d\tilde{t} + d\phi)^2 \\
     &=& - N^2(r) d\tilde{t}^2 + h_{rr}(r) dr^2 +
     h_{\phi\phi}(r) [N^{\phi}(r)d\tilde{t} + d\phi]^2. \nonumber
\end{eqnarray}
Remarkably, the metric now takes on the structure of simple ``diagonal''
ADM's $(2 + 1)$ space-plus-time split form with the lapse, shift functions
and the spatial metric components being given respectively by
\begin{eqnarray}
N^2(r) &=& {\Delta \over \Sigma},   ~~~N^{\phi}(r)=-{a \over \Sigma}, 
\nonumber \\
h_{rr}(r) &=& N^{-2}(r), ~~~h_{\phi\phi}(r) = \Sigma \sin^2 \theta_{0},
~~~h_{r\phi}(r) = h_{\phi r}(r) = 0. \nonumber
\end{eqnarray} 
In terms of this new time coordinate  $\tilde{t}$, therefore, it becomes 
clearer that the $\theta = \theta_{0}$ submanifolds of Kerr-Newman
spacetime have the topology of $R^2 \times S^1$ (which
was not so transparent in the original Boyer-Lindquist time coordinate $t$) 
and hence becomes better-suited for the study of global structure.
That is to say, the global structure of the submanifolds ${\it T}=R^2$
would mirror that of the full $\theta = \theta_{0}$ submanifolds since
each point of  ${\it T}=R^2$ can be thought of as representing $S^1$.
We would like to add a comment here. Of course it is true that it is
the manifold itself that has topology, not the metric. Therefore,
regardless of the metrics one chooses, they all describe the same
manifold with a single topology. However, the point we would like
to make here is that the metric given in new time coordinate $\tilde{t}$
(eq.(4)) demonstrates more clearly that the  $\theta = \theta_{0}$ 
submanifolds it describes has the topology of $R^2 \times S^1$ than
the metric in Boyer-Lindquist time coordinate $t$ (eq.(2)) does. \\
Firstly, consider the $\theta_{0} = 0, ~\pi$ timelike submanifolds 
representing the
``symmetry axis'' of the Kerr-Newman spacetime with the metric being given
by
\begin{eqnarray}
ds^2 = - ({\Delta \over  r^2 + a^2}) dt^2 + ({\Delta \over  r^2 + a^2})^{-1}
dr^2.
\end{eqnarray}
Note that this metric of the symmetry axis is effectively 2-dim.
since it is degenerate along the $\phi$ direction.
And this diagonal, 2-dim. structure of the metric of the
symmetry axis allowed a complete analysis of its global structure as had   
been carried out by Carter$^3$. \\
Secondly, consider the $\theta_{0} = \pi/2$ surface which represents
the ``equatorial plane'' of Kerr-Newman spacetime. The metric of this
submanifold is obtained in the new time coordinate $\tilde{t}$ by setting
 $\theta_{0} = \pi/2$ in eq.(4) 
\begin{eqnarray}
ds^2 = - N^2(r) d\tilde{t}^2 + N^{-2}(r) dr^2 +
r^2 [N^{\phi}(r)d\tilde{t} + d\phi]^2
\end{eqnarray}
with the lapse $N(r)$ and the shift $N^{\phi}(r)$ in above $(2 + 1)$-
split form being given by
\begin{eqnarray}
N^2(r) &=& {\Delta \over r^2} = [ 1 - {2M \over r} + {(a^2 + e^2) \over
r^2}], \nonumber \\
N^{\phi}(r) &=& - {a\over r^2}. \nonumber
\end{eqnarray}
Finally, note that the metric of the  
 $\theta = \theta_{0}$ ($0 < \theta_{0} < \pi/2$) submanifolds given
in eq.(4) are everywhere non-singular including $r =  0$ and possess
exactly the same causal structure (except for the appearance of ergoregion)
as that of the symmetry axis $(\theta = 0)$. Therefore, the maximal analytic
extension of the $\theta = \theta_{0}$ ($0 < \theta_{0} < \pi/2$) 
submanifolds representing their global structure is essentially the
same as that of the symmetry axis first studied by Carter$^3$. 
The metric of the equatorial plane $(\theta_{0} = \pi/2)$ given in eq.(6),
however, possesses a curvature singularity at $r = 0$ as expected (since
it is the ``ring singularity'', $r = 0$, $\theta_{0} = \pi/2$) whereas it exhibits
almost the same causal structure (again except for the presence of the 
ergoregion) as that of the symmetry axis. As a result, the maximal analytic
extension of the equatorial plane is identical to that of the RN 
spacetime$^{1,3}$. Detailed analysis of the maximal analytic extension of the
$\theta =$ const. submanifolds including the transformations to the
Kruskal-type coordinates in which the metric can be cast into the form
\begin{eqnarray}
ds^2 = \Omega^2(r)(-dT^2 + dX^2) + \Sigma \sin^{2}\theta_{0}
[N^{\phi}(r)d\tilde{t} + d\phi]^2
\end{eqnarray}
with $(T, X)$ being the Kruskal-type coordinates and $\Omega(r)$ being
the associated conformal factor
and the exposition of Carter-Penrose conformal
diagrams will be reported in a separate publication. \\
And what makes this type of concrete analysis of the global structure
possible is the fact that in the new time coordinate $\tilde{t}$ given in 
eq.(3), it becomes more apparent that the $\theta =$const. submanifolds 
of Kerr-Newman spacetime with
the metric being given by eq.(4) or (6) has the topology of $R^2 \times
S^1$ which was not so manifest in the old, Boyer-Lindquist time
coordinate $t$. Thus it seems now natural to ask the physical meaning of the 
new time coordinate $\tilde{t}$. 
To get a quick answer to this question, we go back and look at the
coordinate transformation law given in eq.(3) relating the two time
coordinates $t$ and $\tilde{t}$.
Namely, taking the dual of the transformation law
$\delta \tilde{t} = \delta t - (a \sin^2 \theta_{0}) \delta \phi$, 
we get
\begin{eqnarray}
({\partial \over \partial \tilde{t}})^{\mu} &=& ({\partial \over \partial 
t})^{\mu} - {1\over (a \sin^2 \theta_{0})}({\partial \over \partial 
\phi})^{\mu} \nonumber \\
{\rm or} \\
\tilde{\xi}^{\mu} &=& \xi^{\mu} - {1\over (a \sin^2 \theta_{0})} \psi^{\mu}
\nonumber
\end{eqnarray}
where $\xi^{\mu} = (\partial / \partial t)^{\mu}$ and
$\psi^{\mu} = (\partial / \partial \phi)^{\mu}$ denote
Killing fields corresponding to the time translational and the 
rotational isometries of the Kerr-Newman black hole spacetime
respectively and
$\tilde{\xi}^{\mu} = (\partial / \partial \tilde{t})^{\mu}$
denotes the Killing field associated with the isometry under the new
time translation. Now this expression for the new time translational
Killing field  $\tilde{\xi}^{\mu}$ implies that in ``new" time coordinate
$\tilde{t}$, the time translational generator is given by the linear
combination of the old time translational generator and the rotational
generator. In plain English, this means that in ``new" time coordinate,
the action of new time translation consists of the action of old time
translation and the action of rotation in opposite direction to $a$,
i.e., to the rotation direction of the hole. Thus the new time
coordinate $\tilde{t}$ can be interpreted as the coordinate, say,
of a frame which rotates around the axis of the spinning
Kerr-Newman black hole in opposite direction to that of the hole.
Further, by considering the angular velocity,
the angular momentum per unit mass (which will be defined concretely
later on) and the surface gravity at the event horizon of the
hole both in the original Boyer-Lindquist time coordinate $t$ and in the new
time coordinate $\tilde{t}$ and then comparing them, one can explore the
relative physical meaning between $t$ and $\tilde{t}$  in a more 
comprehensive manner. Thus in the following we shall  do this. 
As stated above, since the Kerr-Newman
spacetime is a stationary, axisymmetric solution, it possesses two Killing
fields $\xi^{\mu} = (\partial /\partial t)^{\mu}$ and $\psi^{\mu} =
(\partial /\partial \phi)^{\mu}$ associated with the time-translational
and rotational isometries, respectively.
And it is their linear conbination, $\chi^{\mu} = \xi^{\mu} + \Omega_{H}
\psi^{\mu}$ which is normal to the Killing horizon of the rotating
Kerr-Newman solution. In addition, normally this is the defining equation
of the angular velocity of the event horizon, $\Omega_{H}$$^6$. Thus
from this equation, we first determine the location of the event horizon
and next its angular velocity. Since the Killing
horizon is defined to be a surface on which the Killing field $\chi^{\mu}$
becomes null, in order to find the event horizon, we look for zeros of
$\chi^{\mu}\chi_{\mu} = 0$. A straightforward calculation shows that the
Killing field $\chi^{\mu}$ becomes null at points where
$\Delta = r^2 - 2Mr + a^2 + e^2 = 0$ both in the Boyer-Lindquist time
coordinate, $t$ and in the new time coordinate, $\tilde{t}$. Thus we
have regular inner and outer horizons at 
$r_{\pm} = M \pm \sqrt{M^2 - a^2 - e^2}$ and $r = r_{+}$ is the event
horizon provided $M^2 \geq a^2 + e^2$. Now we are in a position to 
compute the angular velocity at this event horizon. The angular velocity
of the event horizon is given by
\begin{eqnarray}
\Omega_{H} &=& {d\phi \over dt}\vert_{r_{+}} =
-{g_{t\phi}\over g_{\phi \phi}}\vert_{r_{+}} = {a \over r^2_{+} + a^2}, \\
\tilde{\Omega}_{H} &=& {d\phi \over d\tilde{t}}\vert_{r_{+}} =
-{g_{\tilde{t}\phi}\over g_{\phi \phi}}\vert_{r_{+}} = {a \over r^2_{+} + 
a^2 \cos^2 \theta_{0}} ~~~~~(0 \leq \theta_{0} \leq \pi/2) \nonumber
\end{eqnarray}
as measured in the Boyer-Lindquist time coordinate $t$ and in new time
coordinate $\tilde{t}$, respectively. \\
Next, notice that in the Boyer-Lindquist coordinates
(irrespective of choosing $t$ or $\tilde{t}$ as its
time coordinate), the metric component $g_{\phi\phi}$ represents
({\it proper ~distance ~from ~the ~axis ~of ~rotation})$^2$. 
This suggests that the angular velocity we computed above
may be written as
\begin{eqnarray}
\Omega = -{g_{t\phi} \over g_{\phi\phi}} =
{({\it angular ~momentum ~per ~unit ~mass})\over
({\it proper ~distance ~from ~the ~axis ~of ~rotation})^2}\nonumber
\end{eqnarray}
and as a result the ``angular momentum per unit mass" at some
point from the axis of rotation may be identified with
\begin{eqnarray}
J = g_{\phi\phi}\Omega = - g_{t\phi}.
\end{eqnarray}
(As we have seen, the quantity so defined as above
admits clear interpretation of the {\it angular momentum
per unit mass at some point from the axis of rotation}
particularly for $\theta =$const. submanifolds.
But generally, it should be distinguished from the total angular
momentum of the entire Kerr-Newman spacetime measured in the asymptotic 
region, $\hat{J} = (16\pi)^{-1}\int_{S}\epsilon_{\mu\nu\alpha\beta}
\nabla^{\alpha} \psi^{\beta}$ in the notation convention of
ref.6 with $S$ being a large sphere in the asymptotic region
and $\psi^{\mu} = (\partial /\partial \phi)^{\mu}$ being the 
rotational Killing field introduced earlier. Thus in these
definitions, angular momentum per unit mass $J$ may change
under  coordinate transformations although the total angular
momentum $\hat{J}$ remains coordinate-independent.)
Thus at the horizon $r = r_{+}$, the angular momentum per unit mass 
in the usual
Boyer-Lindquist time $t$ and in the new time coordinate $\tilde{t}$ are 
given respectively by  
\begin{eqnarray}
J_{H} &=& \mid g_{t\phi}(r_{+}) \mid = a \sin^2 \theta_{0}
({r^2_{+} + a^2 \over r^2_{+} + a^2 \cos^2 \theta_{0}}), \\
\tilde{J}_{H} &=& \mid g_{\tilde{t}\phi}(r_{+})\mid = a \sin^2 \theta_{0}
~~~~~(0 \leq \theta_{0} \leq \pi/2). \nonumber
\end{eqnarray}
Finally, we turn to the computation of the surface gravity of the 
$\theta =$const. submanifolds of Kerr-Newman black hole.
In physical terms, the surface gravity $\kappa$ is the force that must
be exerted to hold a unit test mass at the horizon and it is given in a
simple formula as$^6$ 
\begin{eqnarray}
\kappa^2 = - {1\over 2} (\nabla^{\mu}\chi^{\nu})(\nabla_{\mu}\chi_{\nu})
\end{eqnarray}
where $\chi^{\mu}$ is as given earlier and the evaluation on the horizon
is understood. And since there are two regular Killing horizons at
$r = r_{+}$ and $r = r_{-}$, we define surface gravities at each of the two
horizons correspondingly. Again a straightforward calculation yields 
\begin{eqnarray}
\kappa_{\pm} &=& {(r_{\pm} - r_{\mp}) \over 2(r^2_{\pm} + a^2)}, \\
\tilde{\kappa}_{\pm} &=& {(r_{\pm} - r_{\mp}) \over 
2(r^2_{\pm} + a^2 \cos^2 \theta_{0})} ~~~~~(0 \leq \theta_{0} \leq \pi/2)
\nonumber
\end{eqnarray}
in Boyer-Lindquist time coordinate $t$ and in new time coordinate $\tilde{t}$
respectively and where we redefined
$\tilde{\kappa}_{\pm} \rightarrow [\sin^2 \theta_{0}(1+\sin^2 \theta_{0})/2]^
{1/2}\tilde{\kappa}_{\pm}$ for $0 < \theta_{0} \leq \pi/2$, but not for
$\theta_{0} = 0$. \\
Thus from eqs. (9), (11) and (13), we can relate quantities with ``tilde"
in new time coordinate $\tilde{t}$ and those in Boyer-Lindquist time 
coordinate $t$ as 
\begin{eqnarray}
\tilde{J}_{H} &=& ({r^2_{+} + a^2 \cos^2 \theta_{0} \over r^2_{+} + a^2}) 
J_{H} < J_{H}, \nonumber \\
\tilde{\Omega}_{H} &=& ({r^2_{+} + a^2 \over r^2_{+} + a^2 \cos^2 \theta_{0}})
\Omega_{H} > \Omega_{H}, \\
\tilde{\kappa}_{\pm} &=& ({r^2_{\pm} + a^2 \over r^2_{\pm} + a^2 \cos^2 
\theta_{0}}) \kappa_{\pm} > \kappa_{\pm}. \nonumber
\end{eqnarray}
Note that in Boyer-Lindquist time coordinate $t$, the angular velocity
$\Omega_{H}$ and the surface gravity  $\kappa_{\pm}$ at the horizon are
independent of the polar angle $\theta = \theta_{0}$ $(0 \leq \theta_{0} \leq
\pi/2)$, i.e., they remain the same for any value of $\theta_{0}$.
In contrast, in the new time coordinate $\tilde{t}$, both
$\tilde{\Omega}_{H}$ and $\tilde{\kappa}_{\pm}$ do have dependence on the
polar angle  $\theta_{0}$ in such a way that they increase with $\theta_{0}$,
i.e., they get minimized at the symmetry axis $(\theta_{0} = 0)$ whereas
get maximized on the equatorial plane $(\theta_{0} = \pi/2)$.
In addition, from eq.(14), note that generally 
$\tilde{\Omega}_{H} > \Omega_{H}$, $\tilde{J}_{H} < J_{H}$ 
and $\tilde{\kappa}_{\pm} > \kappa_{\pm}$
(where the inequalities hold for $\theta_{0} \neq 0$ and up to the redefinition
of $\kappa$). These results indicate that in the ``new" time coordinate
$\tilde{t}$, the $\theta =$const. submanifolds of the Kerr-Newman
black hole has greater angular velocity yet smaller angular momentum per 
unit mass and
greater surface gravity than they do in the usual Boyer-Lindquist time
coordinate $t$. Particularly here, ``possessing greater angular velocity
while smaller angular momentum per unit mass" may first look erroneous. 
But if one
really looks into the details, one can realize that it is no nonsense
since it arises from ``same coordinate distance but different proper
distances" from the axis of rotation to the horizon in the two time 
coordinates $t$ and $\tilde{t}$.
Now from the greater angular velocity, we are led to the conclusion 
that the new time coordinate $\tilde{t}$ defined
by the eq.(3) appears to be the time coordinate, say,  of a frame which 
rotates
around the axis of the spinning Kerr-Newman black hole in opposite direction
to that of the hole with an angular velocity that increases with the polar angle
$\theta_{0}$. Then the smaller angular momentum per unit mass and greater 
surface
gravity can be attributed to the fact that as the angular momentum 
per unit mass decreases when transforming from the Boyer-Lindquist to new 
time coordinate, the surface gravity is expected to increase due to the
effect of centrifugal force. And this conclusion agrees with our
earlier quick interpretation of the ``new" time coordinate $\tilde{t}$.
\\
{\bf IV. Application to Kerr-Newman-type solutions in other gravity theories}
\\
In this section, we shall illustrate that the same type of transformation,
$\tilde{t} = t - (a\sin^2 \theta_{0})\phi $ from the Boyer-Lindquist time
coordinate $t$ to the new time coordinate $\tilde{t}$ allows us to
explore global structure of $\theta =$const. submanifolds of Kerr-Newman-
type solutions found in other gravity theories, most notably low energy
effective string theories.
Thus before we begin, it seems worth mentioning the connection between
black hole physics and string theory. In the study of string theory,
much of the recent attention has been focussed on the construction of
classical solutions such as solitonlike solutions including black hole
solutions. The motivation for such study is the
following. In order to study non-perturbative string theory, one must
include, in addition to the standard Fock space states, the soliton
states in the spectrum. In particular, some of the underlying
symmetries of string theory may become manifest only after including
these solitonic states in the spectrum. As a result, several black
hole solutions have been found as stable extended solitonlike states
in low energy string theory which can be thought of as the string 
theory analogue of important black hole solutions of Einstein gravity.
In the present work, we are particularly interested in rotating,
charged Kerr-Newman-type black hole solutions containing dilaton
and axion corrections in low energy string theory. These solutions 
were obtained first by Sen$^9$ and then in a more general form
(including the Newman-Unti-Tamburino (NUT) parameter) by Gal'tsov
and Kechkin$^{10}$. As it is the case with the Kerr-Newman solution
in Einstein-Maxwell theory, obtaining these rotating, charged
black hole solutions in low energy string theory by directly solving 
classical field equations is highly involved and impractical. 
They were, therefore, obtained by a method for generating new 
solutions from the known ones which can be thought of as a
generalization of Ehlers-Harrison transformations$^8$. \\
First, we begin with the rotating, charged black hole solution
in low energy effective heterotic string theory found by Sen$^9$.
Sen considered the low energy effective theory of heterotic
string described by the action in 4-dim.
\begin{eqnarray}
S = \int d^4x\sqrt{g} e^{-\Phi}( R + \partial_{\mu}\Phi 
\partial^{\mu}\Phi - {1\over 12}H_{\mu\nu\lambda}H^{\mu\nu\lambda}
- {1\over 8}F_{\mu\nu}F^{\mu\nu} )
\end{eqnarray}
where $F_{\mu\nu} = \partial_{\mu}A_{\nu}-\partial_{\nu}A_{\mu}$
is the field strength of the Maxwell field $A_{\mu}$,
$\Phi$ is the dilaton field and 
\begin{eqnarray}
H_{\mu\nu\lambda} = \partial_{\mu}B_{\nu\lambda} 
+  \partial_{\nu}B_{\lambda\mu} +  \partial_{\lambda}B_{\mu\nu}
- [\omega_{3}(A)]_{\mu\nu\lambda} \nonumber
\end{eqnarray}
with $B_{\mu\nu}$ and $[\omega_{3}(A)]_{\mu\nu\lambda}$ being the
antisymmetric tensor gauge field and the gauge Chern-Simons term
respectively (the Lorentz Chern-Simons term has not been included
in the definition of $H_{\mu\nu\lambda}$). And the action above,
which is written in string frame with $g_{\mu\nu}$, can be transformed 
to the usual Einstein conformal frame with $\hat{g}_{\mu\nu}$
(in which the black hole solution will be given later on) via
the conformal transformation
\begin{eqnarray}
\hat{g}_{\mu\nu} = e^{-\Phi} g_{\mu\nu}. \nonumber
\end{eqnarray}
Then, in order to obtain a rotating, charged black hole solution
in this low energy effective string theory, Sen employed the
``twisting procedure"$^9$ that generates inequivalent classical
solutions starting from a given classical solution of string
theory. In particular, using the method$^9$ for generating
charged black hole solution from a charge-neutral solution,
Sen constructed the rotating, charged black hole solution by
starting from a rotating, uncharged black hole solution, i.e.,
the Kerr solution.
In Einstein conformal frame, Sen's metric solution can be cast, 
in terms of Boyer-Lindquist coordinates, to
\begin{eqnarray}
ds^2 = &-& [{\Delta - a^2 \sin^2 \theta \over \Sigma_{s}}] dt^2 -
{2a\sin^2 \theta ([r(r+r_{-}) + a^2] - \Delta) \over \Sigma_{s}} dt d\phi 
\nonumber \\
&+& [{[r(r+r_{-}) + a^2]^2 - \Delta  a^2 \sin^2 \theta \over \Sigma_{s}}]  
\sin^2 \theta d\phi^2 + {\Sigma_{s}\over \Delta} dr^2 + \Sigma_{s} d\theta^2
\end{eqnarray}
where
\begin{eqnarray}
\Sigma_{s} &=& r(r + r_{-}) + a^2 \cos^2 \theta, \nonumber \\
\Delta &=& r^2 - 2Mr + a^2, ~~~r_{-} = 2M\sinh^2 ({\alpha \over 2}).
\nonumber 
\end{eqnarray}
with $\alpha$ being an arbitrary number.
This solution is a 3-parameter family of rotating, charged black hole
solution with the mass $M_{s}$, charge $Q_{s}$, angular momentum
$J_{s}$ and magnetic dipole moment $\mu_{s}$ being given respectively
by
\begin{eqnarray}
M_{s} &=& {M\over 2}(1 + \cosh \alpha), ~~~Q_{s} = {M\over \sqrt{2}}
\sinh \alpha, \nonumber \\
J_{s} &=& {Ma\over 2}(1 + \cosh \alpha), ~~~\mu_{s} = {Ma\over \sqrt{2}}
\sinh \alpha \nonumber
\end{eqnarray}
and with horizons being located at
\begin{eqnarray}
r_{\pm} = M_{s} - {Q^2_{s}\over 2M_{s}} \pm [M^2_{s}(1 - 
{Q^2_{s}\over 2M^2_{s}})^2 - {J^2_{s}\over M^2_{s}}]^{1/2}. \nonumber
\end{eqnarray}
Now, just as we did for the Kerr-Newman solution in Einstein-Maxwell
theory, here we consider the $\theta = \theta_{0}$ $(0 \leq 
\theta_{0} \leq \pi/2)$ timelike submanifolds of Sen's solution above
and next perform the same type of transformation from the
Boyer-Lindquist time coordinate $t$ to the new time coordinate
$\tilde{t}$ as the one given earlier
\begin{eqnarray}
\tilde{t} = t - (a \sin^2 \theta_{0}) \phi \nonumber
\end{eqnarray}
with other spatial coordinates remaining unchanged.
Then in terms of this new time coordinate $\tilde{t}$, the metric of
$\theta = \theta_{0}$ submanifolds again take on the structure of
simple, diagonal, ADM's  $(2 + 1)$ space-plus-time split form
\begin{eqnarray}
ds^2 = -N^2(r)d\tilde{t}^2 + h_{rr}(r)dr^2 + h_{\phi\phi}(r)
[N^{\phi}(r)d\tilde{t} + d\phi]^2 \nonumber
\end{eqnarray}
where the lapse, shift functions and the spatial metric components
are given respectively by
\begin{eqnarray}
N^2(r) &=& {\Delta \over \Sigma_{s}},  ~~~N^{\phi}(r)=-{a \over \Sigma_{s}},
\\
h_{rr}(r) &=& N^{-2}(r), ~~~h_{\phi\phi}(r) = \Sigma_{s} \sin^2 \theta_{0},
~~~h_{r\phi}(r) = h_{\phi r}(r) = 0. \nonumber
\end{eqnarray}
Next, we turn to the rotating, charged black hole solution in 
dilaton-axion gravity found by Gal'tsov and Kechkin$^{10}$. These authors
also considered the low energy effective theory of heterotic string
represented by the action in 4-dim.
\begin{eqnarray}
S = \int d^4x \sqrt{\hat{g}} (\hat{R} + 2\partial_{\mu}\Phi \partial^{\mu}
\Phi + {1\over 2}e^{4\Phi}\partial_{\mu}A \partial^{\mu}A - e^{-2\Phi}
F_{\mu\nu}F^{\mu\nu} - A F_{\mu\nu}\tilde{F}^{\mu\nu})
\end{eqnarray}
where $\tilde{F}^{\mu\nu} = {1\over 2}\epsilon^{\mu\nu\alpha\beta}
F_{\alpha\beta}$ denotes the dual of $F_{\mu\nu}$. Compared to the
action taken by Sen, this action above is written in Einstein conformal
frame with metric $\hat{g}_{\mu\nu}$ and the field strengh of the
Kalb-Ramond field, $H_{\mu\nu\lambda}$ that appeared in Sen's action
is now transformed into the Peccei-Quinn axion field $A$ using the 
dilaton-axion dual symmetry $SL(2,~R)$. Since the axion field appears
explicitly in this format of the theory, the low energy heterotic
string theory considered by Gal'tsov and Kechkin is usually called,
Einstein-Maxwell-Dilaton-Axion (EMDA) theory.
Then, in order to construct a rotating, charged black hole solution
in this EMDA theory, they noted that there are two important symmetries
in bosonic sector of the low energy heterotic string theory which
can be used to generate new classical solutions.
One of them is target space duality $0(d,~d+p)$, which is valid for
EMDA system with $p$ abelian gauge fields whenever variables are
independent of $d$-spacetime coordinates. This is the symmetry that
Sen employed to construct his solution and this solution generation
method is called, ``twisting procedure"$^9$.
The second symmetry is a dilaton-axion (or electric-magnetic) duality
$SL(2,~R)$, which arises in the 4-dim. case for which the field 
strengh of the Kalb-Ramond field, $H_{\mu\nu\lambda}$ can be transformed
into the axion field $A$ as mentioned earlier. It says that a pair
$(\Phi, ~A)$ parametrizes the $SL(2,~R)/SO(2)$ coset.
Gal'tsov and Kechkin took one step further and attempted combining
these two symmetries within a larger group for the case, $p = 1$,
$d = 1$ in 4-dim. And the combined symmetry group turned out to be
larger than the product of the two symmetries. Moreover, its 
nontrivial part generalizes Ehlers-Harrison transformations$^8$ known
in the Einstein-Maxwell theory. The group also contains scale and
gauge transformations. Utilizing these two symmetries, then, they
built a simple way to construct rotating, charged black hole solutions
in EMDA theory from known solutions to vacuum Einstein theory such as
Kerr solution. Thus, in a sense, the solution generation method of
Gal'tsov and Kechkin can be thought of as an extension of twisting
procedure by Sen and their solution (particularly the electrically-
charged solution without magnetic charge and NUT parameter) is
given, in Boyer-Lindquist coordinates, by
\begin{eqnarray}
ds^2 = &-& [{\Delta_{d} - a^2 \sin^2 \theta \over \Sigma_{d}}] dt^2 -
{2a\sin^2 \theta ([r(r+r_{-}) + a^2] - \Delta_{d})\over \Sigma_{d}} dtd\phi
\nonumber \\
&+& [{[r(r+r_{-}) + a^2]^2 - \Delta_{d}a^2 \sin^2 \theta \over \Sigma_{d}}]
\sin^2 \theta d\phi^2 + {\Sigma_{d}\over \Delta_{d}} dr^2 + \Sigma_{d} 
d\theta^2 
\end{eqnarray}
where
\begin{eqnarray}
\Sigma_{d} &=& r(r + r_{-}) + a^2 \cos^2 \theta, \nonumber \\
\Delta_{d} &=& r(r + r_{-}) - 2Mr + a^2, ~~~r_{-} = {Q^2 \over M} 
\nonumber
\end{eqnarray}
and the coordinate ``$r$" here corresponds to $r_{0}$ in the definition of
Gal'tsov and Kechkin$^{10}$. This rotating, charged black hole solution is
characterized by the mass $M$, the electric charge $Q$, and the angular
momentum per unit mass $a$. Its horizons are located at
\begin{eqnarray}
r_{\pm} = M + {Q^2 \over 2M} \pm [M^2(1 - {Q^2\over 2M^2})^2 - a^2]^{1/2}
\nonumber
\end{eqnarray}
and the boundary of the ergosphere, i.e., the static limit is located at
\begin{eqnarray}
r_{\pm} = M + {Q^2 \over 2M} \pm [M^2(1 - {Q^2\over 2M^2})^2 - 
a^2\cos^2 \theta]^{1/2}. \nonumber
\end{eqnarray}
Now, as before, here we consider the $\theta = \theta_{0}$
$(0\leq \theta_{0} \leq \pi/2)$ timelike submanifolds of this rotating,
charged black hole solution in EMDA theory and next perform the
transformation from the Boyer-Lindquist time $t$ to a new time
coordinate $\tilde{t}$
\begin{eqnarray}
\tilde{t} = t - (a\sin^2 \theta_{0}) \phi \nonumber
\end{eqnarray}
as we did before. One can see that, again, in this new time coordinate
$\tilde{t}$, the metric of $\theta = \theta_{0}$ submanifolds take on the 
structure of simple, diagonal ADM's $(2 + 1)$ space-plus-time split
form
\begin{eqnarray}
ds^2 = -N^2(r)d\tilde{t}^2 + h_{rr}(r)dr^2 + h_{\phi\phi}(r)
[N^{\phi}(r)d\tilde{t} + d\phi]^2 \nonumber
\end{eqnarray}
where the lapse, shift functions and the spatial metric components
are given respectively by
\begin{eqnarray}
N^2(r) &=& {\Delta_{d} \over \Sigma_{d}},  ~~~N^{\phi}(r)=-{a \over 
\Sigma_{d}}, \\
h_{rr}(r) &=& N^{-2}(r), ~~~h_{\phi\phi}(r) = \Sigma_{d} \sin^2 \theta_{0},
~~~h_{r\phi}(r) = h_{\phi r}(r) = 0. \nonumber
\end{eqnarray}
Note first that in all the three cases we have considered thus far, i.e.,
the Kerr-Newman solution in Einstein-Maxwell theory, the rotating, charged
black hole solution by Sen and by Gal'tsov and Kechkin in low energy
effective theory of heterotic string, the metric of $\theta =$const.
timelike submanifolds can be cast to a simple, diagonal ADM's
$(2 + 1)$ split form by a ``single", common transformation law for time
coordinate, $\tilde{t} = t - (a\sin^2 \theta_{0})\phi$.
This fact seems to suggests that the transformation law from the 
Boyer-Lindquist time coordinate to the new time coordinate we found, or
equivalently the selection of new time coordinate $\tilde{t}$ was not
accidental, after all. Namely, the coordinate transformation to the frame
which ``rotates around the axis of the spinning hole in opposite direction
to that of the hole with an angular velocity that increases with the 
polar angle $\theta_{0}$" seems to have remarkable physical significance
with general applicability.
Secondly note that, again in terms of this new time coordinate $\tilde{t}$,
it becomes clearer that the $\theta = \theta_{0}$ submanifolds of rotating,
charged black hole spacetime solutions in low energy effective string
theory have the topology of $R^2 \times S^1$. The global structures of the
submanifolds ${\it T} = R^2$, then, would mirror those of the full
$\theta = \theta_{0}$ submanifolds since each point of  ${\it T} = R^2$
can be thought of as representing $S^1$. As we have observed in the
Kerr-Newman spacetime case, now we can conclude that the global structure
(upon the maximal analytic extension) of the $\theta = \theta_{0}$
$(0 \leq \theta_{0} < \pi/2)$ submanifolds is essentially the same as 
that of 
the symmetry axis first studied by Carter$^3$ and the global structure of
the equatorial plane $(\theta_{0} = \pi/2)$ is identical to that of 
the RN spacetime$^3$.
\\
{\bf IV. Discussions}
\\
We now conclude with some comments worth mentioning.
First, it is interesting to note that the metric of $\theta =$const.
submanifolds of Kerr-Newman spacetime expressed in the ``new'' time
coordinate $\tilde{t}$ as given in eq.(4) or (6) takes precisely the
same structure as that of 3-dim. anti-de Sitter ($AdS_{3}$) black hole
solution discovered recently by Banados, Teitelboim and Zanelli (BTZ)$^7$.
This indicates, among other things, that the choice of coordinates in which
BTZ adopted their metric solution ansatz in ref.7 corresponds to the 
``new'' time coordinates $\tilde{t}$ we have introduced in the present work,
{\it not} the usual Boyer-Lindquist time coordinate. Indeed it has been
demonstrated in detail by this author$^{11}$ that by applying the same
``complex coordinate transformation scheme'' as the one Newman et al.$^2$
employed in ``deriving'' Kerr solution from Schwarzschild solution and 
Kerr-Newman solution from RN solution, one can likewise derive rotating
BTZ black hole solution from the nonrotating BTZ black hole solution.
And in doing so the underlying spirit is that a 3-dim. geometry can be 
thought of as the $\theta = \pi/2$-slice of the 4-dim. geometry in which 
one introduces the null tetrad of basis vectors$^2$.
Now, much as the Kerr-Newman solution ``derived'' in this manner naturally
comes in the Boyer-Lindquist coordinates (via the coordinate transformation
from the Kerr coordinates, as is well-known), the rotating BTZ solution
derived in the same manner (i.e., by Newman's complex coordinate 
transformation method) comes in Boyer-Lindquist-type coordinates as 
well$^{11}$.
Then next by performing a transformation to the ``new'' time coordinate,
$\tilde{t} = t - a\phi$ (i.e., $\theta_{0} = \pi/2$- case of the general 
transformation given in eq.(3)), the derived rotating BTZ black hole 
solution finally can be put in the form originally constructed by BTZ.
A merit of this one-to-one correspondence between the $\theta =$const.
submanifolds of Kerr-Newman black hole and the rotating BTZ solution 
is that now one can carry out the maximal analytic extension of the 
$\theta =$const. submanifolds of Kerr-Newman spacetime following a similar
procedure taken for the complete study of the global structure of rotating
BTZ black hole solution which was possible since it has been done in the
new time coordinate$^7$ in which the metric takes much simpler form and
exhibits more clearly that the spacetime has the topology 
of $R^2 \times S^1$. \\
Next, we would like to point out the complementary roles played
by the two alternative time coordinates $t$ and $\tilde{t}$.
The Boyer-Lindquist time coordinate $t$ is the usual Killing time
coordinate with which one can obtain the rotating hole's characteristics
such as the angular velocity of the event horizon or the surface gravity
as measured by an outside observer who is ``static" with respect to,
say, a distant star. The ``new" time coordinate $\tilde{t}$, on the
other hand, is a kind of unusual in that it can be identified with the 
time coordinate of a non-static frame which rotates around the axis of
the Kerr-Newman black hole in opposite direction to that of the hole 	
with an angular velocity that increases with the polar angle.
This new time coordinate, however, is particularly advantageous in
exploring the global stucture of the $\theta =$ const. submanifolds of 
Kerr-Newman black hole since
it allows one to transform to Kruskal-type coordinates and hence eventually
allows one to draw the Carter-Penrose diagrams much more easily than the 
case when one employs the usual Boyer-Lindquist time coordinate.

\vspace{2cm}

{\bf \large References}

\begin{description}

\item {1} S. W. Hawking and G. F. R. Ellis, {\it The Large Scale Structure of Space-Time}
(Cambridge University Press, 1973).  
\item {2} R. P. Kerr, Phys. Rev. Lett. {\bf 11}, 237 (1963) ;
E. T. Newman and A. I. Janis, J. Math. Phys. {\bf 6}, 915 (1965) ;
E. T. Newman, E. Couch, R. Chinnapared, A Exton, A. Prakash, and R. Torrence,
J. Math. Phys. {\bf 6}, 918 (1965).
\item {3} J. C. Graves and D. R. Brill, Phys. Rev. {\bf 120}, 1507 (1960) ;
B. Carter, Phys. Rev. {\bf 141}, 1242 (1966) ; Phys. Rev. {\bf 174} (1968) ;
R. H. Boyer and R. W. Lindquist, J. Math. Phys. {\bf 8}, 265 (1967). 
\item {4} D. Finkelstein, Phys. Rev. {\bf 110}, 965 (1958).
\item {5} M. D. Kruskal, Phys. Rev. {\bf 119}, 1743 (1960).
\item {6} R. M. Wald, {\it General Relativity} (Univ. of Chicago Press, Chicago, 1984).
\item {7} M. Banados, C. Teitelboim, and J. Zanelli, Phys. Rev. Lett. {\bf 69}, 1849
(1992) ;  M. Banados, M. Henneaux, C. Teitelboim, and J. Zanelli, Phys. Rev. {\bf D48},
1506 (1993).
\item {8} J. Ehlers, in {\it Les Theories Relativistes de la Gravitation},
P275 (CNRS, Paris, 1959) ; B. K. Harrison, J. Math. Phys. {\bf 9}, 1774 
(1968). 
\item {9} A. Sen, Phys. Rev. Lett, {\bf 69}, 1006 (1992) ; S. Hassan and
A. Sen, Nucl. Phys. {\bf B375}, 103 (1992) ; A. Sen, Nucl. Phys. {\bf B388},
475 (1992).
\item {10} D. V. Gal'tsov and O. V. Kechkin, Phys. Rev. {\bf D50}, 7394 
(1994) ; A. Ya. Burinskii, Phys. Rev. {\bf D52}, 5826 (1995).
\item {11} H. Kim, will be published elsewhere.

\end{description}

\end{document}